\documentclass[aps, prb, showpacs, showkeys, preprint, eqsecnum, floatfix, superscriptaddress, nofootinbib]{revtex4-1}

\usepackage{hyperref}
\usepackage{amssymb}
\usepackage{amsmath}
\usepackage{amsthm}
\usepackage{bm}
\usepackage{calrsfs}
\usepackage{color}
\usepackage{graphicx}
\usepackage[normalem]{ulem} 

\newcommand{\bq}{\begin{eqnarray}}
\newcommand{\eq}{\end{eqnarray}}
\newcommand{\bqn}{\begin{eqnarray*}}
\newcommand{\eqn}{\end{eqnarray*}}

\newcommand{\rr}{\mathbf{r}}

\newcommand{\nablab}{\pmb{\nabla}}

\newcommand{\calp}{{\cal P}}
\newcommand{\calh}{{\cal H}}

\newcommand{\cala}{{\cal A}}
\newcommand{\calo}{{\cal O}}

\begin{document}
%%%%%%%%%%%%%%%%%%%%%%%%%%%%%%%%%%%%%%%%%%%%%%%%%%%%%%%%%%%%%%%%%%%%%%%%%%%%%%
%%%%%%%%%%%%%%%%%%%%%%%%%%%%%%%%%%%%%%%%%%%%%%%%%%%%%%%%%%%%%%%%%%%%%%%%%%%%%%
%%%%%%%%%%%%%%%%%%%%%%%%%%%%%%%%%%%%%%%%%%%%%%%%%%%%%%%%%%%%%%%%%%%%%%%%%%%%%%
\title{Jellium at finite temperature using the restricted worm algorithm} 

\author{Riccardo Fantoni}
\email{riccardo.fantoni@posta.istruzione.it}
\affiliation{Universit\`a di Trieste, Dipartimento di Fisica, strada
  Costiera 11, 34151 Grignano (Trieste), Italy}
\date{\today}

\begin{abstract}
We study the Jellium model of Wigner at finite, non-zero, temperature 
through a computer simulation using the canonical path integral worm algorithm 
where we successfully implemented the fixed-nodes free particles restriction 
necessary to circumvent the fermion sign problem. Our results show good agreement 
with the recent simulation data of Brown et al. and of other similar computer 
experiments on the Jellium model at high density and low temperature. Our 
algorithm can be used to treat any quantum fluid model of fermions at finite, 
non zero, temperature and has never been used before in literature. 
\end{abstract}



\maketitle
%%%%%%%%%%%%%%%%%%%%%%%%%%%%%%%%%%%%%%%%%%%%%%%%%%%%%%%%%%%%%%%%%%%%%%%%%%%%%%
\section{Introduction}
%%%%%%%%%%%%%%%%%%%%%%%%%%%%%%%%%%%%%%%%%%%%%%%%%%%%%%%%%%%%%%%%%%%%%%%%%%%%%%
\label{sec:introduction}

The free electron gas or the {\sl Jellium} model of Wigner
\cite{Fantoni2013} is the simplest physical model for the valence
electrons in a metal \cite{Ashcroft-Mermin} (more generally it is an
essential ingredient for the study of ionic liquids (see
Ref. \cite{Hansen} Chapter 10 and 11): molten-salts, liquid-metals,
and ionic-solutions) or the plasma in the 
interior of a white dwarf \cite{Shapiro-Teukolsky}. It can be imagined
as a system of pointwise electrons of charge $e$ made 
thermodynamically stable by the presence of a uniform, inert,
neutralizing background of opposite charge density inside which they
move. In this work we will only be interested in Jellium in
three dimensional Euclidean space even if some progress has been made
to study this system in curved surfaces, too.
\cite{Fantoni03jsp,Fantoni2008,Fantoni2012,Fantoni2012b,Fantoni18a}

The zero temperature, ground-state, properties of the statistical
mechanical Jellium model thus depends just on the electronic density $n$, or
the Wigner-Seitz radius $r_s=(3/4\pi n)^{1/3}/a_0$ where $a_0$ is Bohr
radius, or the Coulomb coupling parameter $\Gamma=e^2/(a_0r_s)$. Free
electrons in metallic elements \cite{Ashcroft-Mermin} has $2\lesssim
r_s \lesssim 4$, whereas in the interior of a white dwarf
\cite{Shapiro-Teukolsky} $r_s\simeq 0.01$. This model has been 
intensively studied in the second half of last century.

The finite, non-zero, temperature model depends additionally on a parameter 
$\Theta=T/T_F$ where $T$ is the absolute temperature and $T_F$ the Fermi 
temperature. This model has received much attention more recently.

The past two decades have witnessed an impressive progress in
experiments and also in quantum Monte Carlo simulations, which have
provided the field with the most accurate thermodynamic data
available. The simulations started with the pioneering work by
Ceperley and co-workers later developed by Filinov and
co-workers. These has been carried on for the pure Jellium model
\cite{Brown2013,Brown2014,Schoof2011,Schoof2015,Dornheim2015,Dornheim2016,Groth2017,Malone2016,Filinov2015},
for hydrogen, hydrogen-helium mixtures, and electron-hole plasmas. Also, 
we recently applied our newly developed simulation methods to the 
one-component system of charged bosons and fermions, both in the three 
dimensional Euclidean space and on the surface of a sphere, and to the 
binary fermion-boson plasma mixture at finite temperature
\cite{Fantoni18a,Fantoni18b}. In the latter study, we discussed the
thermodynamic stability, from the simulation point of view, of the
two-component mixture where the two species are both bosons, both
fermions, and one boson and one fermion. Shortly after our results
were published other groups reported \cite{Dornheim2018} about computer
experiments using methods partly similar to ours.

Today we are able to simulate on a computer the structural and
thermodynamic properties of Jellium at finite, non zero,
temperature. This allows us to predict thermodynamic states that would
be rather difficult to obtain in nature or in the laboratory, such as
Jellium under extreme conditions, partially polarized Jellium, etc.. In
this work we will carry on some of these path integral simulations
which make use of the Monte Carlo technique. Monte Carlo is the best known
method to compute a path integral. \cite{Ceperley1995} The {\sl computer
experiment} is alternative to theoretical analytic
approximations like the Random-Phase-Approximation.
\cite{Hansen1973,Hansen1975,Gupta1980,Perrot1984,Singwi1968,Tanaka1986,Perrot2000,Dharma2000}

As will be made clear in Section \ref{sec:sim}, untill recently, we were
unable to obtain exact numerical results even through computer
experiments, since one had to face the so called {\sl fermions sign 
  problem} which had not been solved before the advent of recent
simulation techniques \cite{Groth2017,Dornheim2016}. When it was demonstrated 
that the fermions sign problem can be partly avoided and nearly exact results
for the thermodynamic functions can be obtained with an error below 1\%,. In other 
words, we were not able to extract exact results not
even numerically from a simulation for fermions, unlike for bosons or
boltzmannons. Therefore, in order to circumvent the fermion sign
problem, we will here resort to the most widely used approximation in
quantum Monte Carlo that is the {\sl restricted path integral} 
fixed-nodes method. \cite{Ceperley1991,Ceperley1996} But unlike previous
studies we will implement this method upon the {\sl worm} algorithm
\cite{Prokofev1998,Boninsegni2006a} in the canonical
ensemble. Recently, we carried on \cite{Fantoni18c} simulations in the
grand canonical ensemble; in the present study we will instead worry about a
precise comparison with the data of Brown {\sl et al.} \cite{Brown2013} 
who worked in the canonical ensemble. The worm algorithm is preferable over 
the usual path integral Monte Carlo methods \cite{Ceperley1995} since it is 
able to build the sum over the permutation through a menu of moves on open 
paths---the worms---instead of sampling the permutation sum explicitly.

The work is organized as follows: in Section \ref{sec:model} we
describe the Jellium model from a statistical physics point of view, in 
Section \ref{sec:sim} we describe the simulation method, in Section 
\ref{sec:problem} we outline the problem we want to solve on the computer, 
in Section \ref{sec:algorithm} we presents our new algorithm in detail,
Section \ref{sec:results} is for our numerical results, and in Section 
\ref{sec:conclusions} we summarize our concluding remarks.

%%%%%%%%%%%%%%%%%%%%%%%%%%%%%%%%%%%%%%%%%%%%%%%%%%%%%%%%%%%%%%%%%%%%%%%%%%%%%%
\section{The model}
%%%%%%%%%%%%%%%%%%%%%%%%%%%%%%%%%%%%%%%%%%%%%%%%%%%%%%%%%%%%%%%%%%%%%%%%%%%%%%
\label{sec:model}

The {\sl Jellium} model of Wigner \cite{March-Tosi, Singwi1981,
  Ichimaru1982, Martin88} is an assembly of $N_+$ spin up pointwise
electrons and $N_-$ spin down pointwise electrons of charge $e$ moving
in a positive, inert background that ensures charge neutrality. The
total number of electrons is $N=N_++N_-$ and the average particle
number density is $n=N/\Omega$, where $\Omega$ is the volume of the
electron fluid. In the volume $\Omega=L^3$ there is a uniform, neutralizing
background with a charge density $\rho_b=-en$. So that the total
charge of the system is zero. The fluid polarization is then
$\xi=|N_+-N_-|/N$: $\xi=0$ in the unpolarized (paramagnetic) case and
$\xi=1$ in the fully polarized (ferromagnetic) case.

Setting lengths in units of $a=(4\pi n/3)^{-1/3}$ and energies in
Rydberg's units, $\text{Ry}=\hbar^2/2ma_0^2$, where $m$ is the electron
mass and $a_0=\hbar^2/me^2$ is the Bohr radius, the Hamiltonian of
Jellium is 
\bq 
\calh&=&-\frac{1}{r_s^2}\sum_{i=1}^N\nablab_{\rr_i}^2+V(R)~,\\ \label{p-e}
V&=&\frac{1}{r_s}\left(2\sum_{i<j}\frac{1}{|\rr_i-\rr_j|}+
\sum_{i=1}^Nr_i^2+
v_0\right)~,
\eq
where $R=\{\rr_1,\rr_2,\ldots,\rr_N\}$ with $\rr_i$ the
coordinate of the $i$th electron, $r_s=a/a_0$, 
and $v_0$ a constant containing the self energy of the background.
Note that the presence of the neutralizing background produces the
harmonic confinement shown in Eq. (\ref{p-e}).

The kinetic energy scales as $1/r_s^2$ and the potential
energy (particle-particle, particle-background, and
background-background interaction) scales as $1/r_s$, so for small
$r_s$ (high electronic densities), the kinetic energy dominates and
the electrons behave like an ideal gas. In the limit of large $r_s$,
the potential energy dominates and the electrons crystallize into a
Wigner crystal. \cite{Wigner1934} No liquid phase is realizable within
this model since the pair-potential has no attractive parts, even
though a superconducting state \cite{Leggett1975} may still be
possible (see chapter 8.9 of Ref. \cite{Giuliani-Vignale} and
Ref. \cite{Pollock1987}).

The Jellium in its ground-state has been solved either by integral 
equation theories \cite{Singwi1968} or by computer experiments
\cite{Ceperley1980} in the second half of last century but more recently 
it has been studied at finite, non-zero, temperatures by several research 
groups.  
\cite{Brown2013,Brown2014,Schoof2011,Dornheim2015,Dornheim2016,Groth2017,Malone2016,Filinov2015} 

following Brown {\sl et al.} \cite{Brown2013}, it is convenient to introduce
the {\sl electron degeneracy parameter} $\Theta=T/T_F$ for the Jellium at finite 
temperature, where $T_F$ is the Fermi temperature of either the unpolarized ($\xi=0$) 
or polarized ($\xi=1$) system 
\bq
T_F=T_D\frac{(2\pi)^2}{2[(2-\xi)\alpha_3]^{2/3}},
\eq
$\xi$ is the polarization of the fluid, $\alpha_3=4\pi/3$ is the volume of 
the unit sphere, and 
\bq
T_D=\frac{n^{2/3}\hbar^2}{mk_B}=
\frac{\hbar^2}{mk_B\alpha_3^{2/3}(a_0r_s)^2},
\eq
is the degeneracy temperature \cite{Ceperley1995}, {\sl i.e.} the temperature at 
which the de Broglie thermal wavelength becomes comparable to the mean 
separation between the particles ($\propto n^{-1/3}$). For temperatures
higher than $T_D$ quantum effects are less relevant.

The state of the fluid will also depend upon the {\sl Coulomb
coupling parameter}, $\Gamma=e^2/(a_0r_s)k_BT$ \cite{Brown2013}, so that 
\bq
\Theta=\frac{r_s}{\Gamma}\left[\frac{2(2-\xi)^{2/3}\alpha_3^{4/3}}
{(2\pi)^2}\right].
\eq 

The behavior of the internal energy of Jellium in its ground-state
($\Theta=0$) has been determined through Diffusion Monte Carlo (DMC)
by Ceperley and Alder. \cite{Ceperley1980} Three phases of the fluid
appeared, for $r_s<75$ the stable phase is the one of the unpolarized
Jellium, for $75<r_s<100$ the one of the polarized fluid, and for
$r_s>100$ the one of the Wigner crystal. They used systems from $N=38$
to $N=246$ electrons. 

It was shown in Ref. \cite{Schoof2015} that the data of Brown {\sl et
  al.} \cite{Brown2013,Brown2014}, for the finite, non-zero temperature 
case, are incaccurate at high densities, $r_s\lesssim 1$. This appears to be a 
systematic error, of up to 10\%, of the restricted path integral fixed node 
method. Thus, it would be interesting to know whether this problem may be solved 
with our present method, which seems a promising route to access higher 
densities. They provide results for the thermodynamic
properties of Jellium with 33 fully polarized, $\xi=1$ electrons and 66
unpolarized, $\xi=0$ electrons, in the warm-dense regime:
$r_s=1,2,4,6,8,10,40$ and $\Theta=0.0625,0.125,0.25,0.5,1,2,4,8$. 

%%%%%%%%%%%%%%%%%%%%%%%%%%%%%%%%%%%%%%%%%%%%%%%%%%%%%%%%%%%%%%%%%%%%%%%%%%%%%%
\section{The simulation} 
%%%%%%%%%%%%%%%%%%%%%%%%%%%%%%%%%%%%%%%%%%%%%%%%%%%%%%%%%%%%%%%%%%%%%%%%%%%%%%
\label{sec:sim}

The {\sl density matrix} of a system of many fermions at temperature
$k_BT=\beta^{-1}$ can be written as an integral over all paths 
$\{R_t\,|\,0\le t\le\beta\}$ 
\bq
\rho_F(R_\beta,R_0;\beta)=\frac{1}{N!}\sum_\calp(-1)^\calp\oint_{\calp
  R_0\to R_\beta}dR_t\,\exp(-S[R_t]),
\eq
where $R_t=\{\rr_1(t),\ldots,\rr_N(t)\}$ represents the positions of all the 
particles at {\sl imaginary time} $t$. The path begins at $\calp R_0$ and 
ends at $R_\beta$; $\calp$ is a permutation of particles labels. For 
non-relativistic particles interacting with a potential $V(R)$, the {\sl action} of 
the path, $S[R_t]$, is given by
\bq
S[R_t]=\int_0^\beta dt\,\left[\frac{r_s^2}{4}\left|\frac{dR_t}{dt}\right|^2
+V(R_t)\right].
\eq 
Thermodynamic properties, such as the energy, are related to the
diagonal part of the density matrix, so that the path returns to its
starting place or to its permutation $\calp$ after a time $\beta$.

To perform Monte Carlo calculations of the integrand, one makes the
imaginary time discrete with a {\sl time step} $\tau$, so that one has
a finite (and hopefully small) number of time slices and thus an
isomorphic classical system of $N$ particles in $M=\beta/\tau$ time
slices; an equivalent $NM$ particle classical system of ``polymers''.
\cite{Ceperley1995} 

Note that in addition to sampling the path, the permutation is also
sampled. This is equivalent to allowing the ring polymers to connect
in different ways. This macroscopic ``percolation'' of the polymers is
directly related to superfluidity, as Feynman
\cite{Feynman1953a,Feynman1953b,Feynman1953c} first showed for bosons. Any
permutation can be broken into cycles. Superfluid behavior can occur
at low temperature when the probability of exchange cycles on the
order of the system size is non-negligible. The {\sl superfluid fraction} 
can be computed in a path integral Monte Carlo (PIMC) calculation as
described in Ref. \cite{Pollock1987}. The same method could be used to
calculate the {\sl superconducting fraction} in Jellium at low
temperature. However, the straightforward application of those
techniques to Fermi systems means that odd permutations must be subtracted 
from the integrand. This is the ``fermions sign problem''
\cite{Ceperley1991} first noted by Feynman \cite{Feynman-Hibbs} who
after describing the path integral theory for boson superfluid $^4$He,
pointed out: 
``{\sl The} [path integral] {\sl expression for Fermi particles, such
as $^3$He, is also easily written down. However in the case of
liquid $^3$He, the effect of the potential is very hard to
evaluate quantitatively in an accurate manner. The reason for this is
that the contribution of a cycle to the sum over permutations is
either positive or negative depending whether the cycle has an odd or
an even number of atoms in its length [$\ldots$]. At very low temperature 
[$\ldots$] it is very difficult to sum an alternating series of large terms 
which are decreasing slowly in magnitude when a precise analytic formula for
each term is not available.}''

Thermodynamic properties are averages over the thermal, $N$-fermions
density matrix which is defined as a thermal occupation of the exact
eigenstates $\phi_i(R)$
\bq
\rho_F(R,R';\beta)=\sum_i\phi_i^*(R)e^{-\beta E_i}\phi_i(R').
\eq
The partition function is the trace of the density matrix
\bq
Z(\beta)=e^{-\beta F}=\int dR\,\rho_F(R,R;\beta)=\sum_ie^{-\beta E_i}.
\eq
Other thermodynamic averages are obtained as
\bq
\langle\calo\rangle=Z(\beta)^{-1}\int dR dR'\,\langle R|\calo|R'\rangle
\rho_F(R',R;\beta).
\eq

Note that for any density matrix the diagonal part is always positive
\bq
\rho_F(R,R;\beta)\ge 0,
\eq
so that $Z^{-1}\rho_F(R,R;\beta)$ is a proper probability
distribution. It is the diagonal part which we need for many
observables, so that probabilistic ways of calculating those
observables are, in principle, possible.

Path integrals are constructed using the product property of density
matrices
\bq
\rho_F(R_2,R_0;\beta_1+\beta_2)=\int dR_1\,
\rho_F(R_2,R_1;\beta_2)\rho_F(R_1,R_0;\beta_1),
\eq
which holds for any sort of density matrix. If the product property is
used $M$ times we can relate the density matrix at a temperature
$\beta^{-1}$ to the density matrix at a temperature $M\beta^{-1}$. The
sequence of intermediate points $\{R_1,R_2,\ldots,R_{M-1}\}$ is the
path, and the {\sl time step} is $\tau=\beta/M$. As the time step gets
sufficiently small the Trotter theorem tells us that we can assume
that the kinetic ${\cal T}$ and potential ${\cal V}$ operator commute
so that: $e^{-\tau\calh}=e^{-\tau{\cal T}}e^{-\tau{\cal V}}$ and the
{\sl primitive approximation} for the fermions density matrix is
found. \cite{Ceperley1995} The Feynman-Kac formula for the
fermions density matrix results from taking the limit $M\to\infty$.
The price we have to pay for having an explicit expression for the
density matrix is additional integrations; all together
$3N(M-1)$. Without techniques for multidimensional integration,
nothing would have been gained by expanding the density matrix into a
path. Fortunately, simulation methods can accurately treat such
integrands. It is feasible to make $M$ rather large, say in the
hundreds or thousands, and thereby systematically reduce the time-step
error.

One can then measure \cite{Ceperley1995} the internal energy (kinetic plus 
potential energy) per particle using the thermodynamic estimator, the pressure 
using the virial theorem estimator, the static structure (the radial 
distribution function), and the superconducting fraction of Jellium. 

One solution to Feynman's task of rearranging terms to keep only
positive contributing paths for diagonal expectation values is the
restricted or fixed-nodes path integral identity. Suppose $\rho_F$ is
the density matrix corresponding to some set of quantum numbers which
is obtained by using the antisymmetrization operator $\cala$ acting on 
the same spin groups of particles on the distinguishable particle density 
matrix. Then the following {\sl Restricted Path Integral} identity holds
\cite{Ceperley1991,Ceperley1996} 
\bq \label{rpii}
\rho_F(R_\beta,R_0;\beta)=\int dR'\,\rho_F(R',R_0;0)
\oint_{R'\to R_\beta\in\gamma(R_0)}dR_t\,e^{-S[R_t]},
\eq
where the subscript means that we restrict the path integration to paths 
starting at $R'$, ending at $R_\beta$ and node-avoiding (those for which 
$\rho_F(R_t,R_0;t)\neq 0$ for all $0<t\le\beta$), {\sl i.e.} paths staying 
inside the {\sl reach} of the reference point $R_0$, \cite{Ceperley1996} $
\gamma(R_0)$ or the {\sl nodal cell} \cite{Ceperley1991}. The weight of the 
walk is $\rho_F(R',R_0;0)=(N!)^{-1}\sum_\calp(-)^\calp\delta(R'-\calp R_0)$. It 
is clear that the contribution of all the paths for a single element of the 
density matrix will be of the same sign, thus solving the sign problem;
positive if $\rho_F(R',R_0;0)>0$, negative otherwise. On the diagonal the 
density matrix is positive and on the path restriction we can always choose
$\rho_F(R_t,R_0;t)>0$ for $0<t\le\beta$, then only even permutations are allowed 
since $\rho_F(R_0,\calp R_0;\beta)=(-)^\calp\rho_F(R_0,R_0;\beta)$. It is then
possible to use a bosons calculation to get the fermions case once the 
restriction has been correctly implemented. 

The problem we now face is that the unknown density matrix appears
both on the left-hand side and on the right-hand side of
Eq. (\ref{rpii}) since it is used to define the criterion of
node-avoiding paths. To apply the formula directly, we would somehow
have to self-consistently determine the density matrix. In practice
what we need to do is make an {\sl ansatz}, which we call $\rho_T$,
for the nodes of the density matrix needed for the restriction. The
{\sl trial density matrix}, $\rho_T$, is used to define the trial reach: 
$\gamma_T(R_0)$. 

Then if we know the reach of the fermion density matrix we can use the
Monte Carlo method to solve the fermion problem, restricting the path
integral (RPIMC) to the space-time domain where the density matrix has a
definite sign (this can be done, for example, using a trial density
matrix whose nodes approximate well the ones of the true density
matrix). Furthermore, we use the antisymmetrization operator to extend 
it to the whole configuration space (using the {\sl tiling} \cite{Ceperley1991} 
property of the reach), $\bigcup_{\calp_e}\gamma_T(\calp_e R_0)$, where only even 
permutations $\calp_e$ are needed. This 
will require the complicated task of sampling the permutation space of the $N$-
particles. \cite{Ceperley1995} Recently, an intelligent method has been devised  
to perform this sampling through a new algorithm called the
{\sl worm} algorithm. \cite{Prokofev1998,Boninsegni2006a} In order to sample the
path in coordinate space, one generally uses various generalizations of
the Metropolis rejection algorithm \cite{Metropolis} and the {\sl
  bisection method} \cite{Ceperley1995} in order to accomplish
multislice moves which becomes necessary as $\tau$ decreases.  

The {\sl pair-product approximation} for the action \cite{Ceperley1995} was 
used by Brown {\sl et al.} \cite{Brown2013} to
write the many-body density matrix as a product of high-temperature,
two-body density matrices. \cite{Ceperley1995} The pair Coulomb
density matrix was determined using the results of Pollock
\cite{Pollock1988}, even if these could be improved using the results
of Vieillefosse. \cite{Vieillefosse1994,Vieillefosse1995} This
procedure comes with an error that scales as $\sim\tau^3/r_s^2$ where
$\tau=\beta/M$ is the {\sl time step}, with $M$ the number of
imaginary time discretizations. A more dominate form of time step error
originates from paths which cross the nodal constraint in
a time less than $\tau$. To help alleviate this effect, Brown {\sl et
  al.} \cite{Brown2013} use an image action to discourage paths from
getting too close to nodes. Additional sources of error are the finite
size one and the sampling error of the Monte Carlo procedure
itself. In their analysis, for the highest density points, statistical 
errors are an order of magnitude higher than time step errors. 

In our calculation, for simplicity, we will use the {\sl primitive approximation} 
\cite{Ceperley1995} for the action. This procedure comes with an error that 
scales as $\sim\tau^2/r_s^2$. And we will have the additional sources of 
error due to the finite size and the sampling of the Monte Carlo procedure
itself, as usual. For the highest density points, statistical errors are of order 
$10^{-3}$, in the potential energy or in the pressure, whereas 
$\tau^2/r_s^2\approx 10^{-6}$.

%%%%%%%%%%%%%%%%%%%%%%%%%%%%%%%%%%%%%%%%%%%%%%%%%%%%%%%%%%%%%%%%%%%%%%%%%%%%%%
\section{The problem}
%%%%%%%%%%%%%%%%%%%%%%%%%%%%%%%%%%%%%%%%%%%%%%%%%%%%%%%%%%%%%%%%%%%%%%%%%%%%%%
\label{sec:problem}

Like Brown {\sl et al.} \cite{Brown2013} we adopted as trial density matrix for 
the path integral nodal restriction a free fermion density matrix. This allowed 
us to implement the restriction in the path integral calculation from the worm
algorithm \cite{Boninsegni2006a,Boninsegni2006b} to the reach of the
reference point in the moves ending in the Z sector: remove, close,
wiggle, and displace. The worm algorithm is a particular path integral
algorithm where the permutations need not to be sampled as they are
generated with the simulation evolution. Instead of the pair-product action used 
by Brown {\sl et al.} \cite{Brown2013}, we used the {\sl primitive
approximation} for the action \cite{Ceperley1995} and modified the original worm 
algorithm so that it would work in the presence of the nodal restriction and in a 
canonical ensemble calculation at fixed number of particles $N$, volume 
$\Omega=N\alpha_3$, and temperature $T$. We should mention that, due to the 
choice of approximation for the action, our results will suffer of some 
additional systematic error respect to the data of Brown {\sl et al.}, although
small.

The restriction implementation is rather simple: we just reject the
move whenever the proposed path is such that the ideal fermion density
matrix calculated between the reference point and any of the time
slices subject to newly generated particles positions has a negative
value. Our algorithm is described in detail in the following section.

The trial density matrix used to perform the restriction of the 
fixed-nodes path integral is chosen as the one of ideal fermions which is
given by 
\bq \label{ifdm}
\rho_0(R,R';t)\propto\cala \left[e^{
-\frac{(\rr_i-\rr_j')^2}{4\lambda t}}\right]\stackrel{\xi=1}{=}
\det\left[\exp\left(
-\frac{r_s^2(\rr_i-\rr_j')^2}{4t}\right)\right],
\eq
where $\lambda=\hbar^2/2m$, $t$ is the imaginary time, and $\cala$ is the 
antisymmetrization
operator acting on the same spin groups of particles, which for polarized 
electrons reduces to a single determinant, and the distances $\sqrt{(\rr_i-
\rr_j')^2}$ are calculated taking care, as usual, of the wrapping due to the 
periodic boundary conditions. We expect this approximation to be best at high 
temperatures (high $\Theta$) and high densities (low $r_s$) when
the quantum and correlation effects are weak. Clearly in a simulation of the 
ideal gas ($V=0$) this restriction returns the exact result for fermions.

The Coulomb potential is treated through the method of Fraser {\sl et
  al.} \cite{Fraser1996} which is alternative to the Ewald summation of Natoli 
and Ceperley \cite{Natoli1995}, to cure its long-range nature.

%%%%%%%%%%%%%%%%%%%%%%%%%%%%%%%%%%%%%%%%%%%%%%%%%%%%%%%%%%%%%%%%%%%%%%%%%%%%%%
\section{Our algorithms} 
%%%%%%%%%%%%%%%%%%%%%%%%%%%%%%%%%%%%%%%%%%%%%%%%%%%%%%%%%%%%%%%%%%%%%%%%%%%%%%
\label{sec:algorithm}

Our algorithm, that we will call algorithm A, briefly presented in the previous 
section is based on the worm algorithm of Boninsegni {\sl et al.}
\cite{Boninsegni2006a,Boninsegni2006b,Fantoni2014b,Fantoni15b,Fantoni16a}. 
The algorithm of Boninsegni {\sl et al.} solves the path integral in the grand canonical ensemble and 
uses a menu of 9 moves. Three are self-complementary: swap, displace, and wiggle, and 
the other six are 3-couples of complementary moves: insert-remove,
open-close, and advance-recede. These moves act on ``worms'' with an
head {\sl Ira} and a tail {\sl Masha} in the $\beta$-periodic imaginary 
thermal time, which can swap a portion of their bodies (swap move), can move 
forward and backward (advance-recede moves), can be subdivided in two
or joined into a bigger one (open-close moves), and can be born or die
(insert-remove moves) since we are working in the grand-canonical
ensemble. The configuration space of the worms is called the G sector. When 
the worms recombine to form a closed path (``world line'') we enter the so 
called Z sector and the path can translate in space (displace move) and can 
propagate in space through the bisection algorithm (wiggle move), carefully 
explained in Ref. \cite{Ceperley1995}. In order to reduce the grand canonical 
algorithm to a canonical calculation it is sufficient to choose the chemical 
potential equal to zero everywhere in the algorithm and to reject all the moves 
attempting to change the number of particles $N$ in the Z sector. Of course it 
is necessary to initialize the calculation from a path containing the given 
number $N$ of particles. 

In order to get the restricted path integral we choose the trial density matrix 
as the one of the non-interacting fermions (\ref{ifdm}) and restrict the Z to Z 
and the G to Z moves, that is: displace, wiggle, close, and remove. In order 
to implement the restriction we reject the move whenever the proposed path is 
such that the ideal fermions density matrix of Eq. (\ref{ifdm}) calculated 
between the reference point $R_0$ and any of the time slices subject to newly 
generated particles positions, $R_t$ with $0<t\leq\beta$, changes sign. That is, 
whenever the path ends up in a region not belonging to the trial reach of the 
reference point. So, we implemented the rejection every time we encounter $
\rho_0(R_t,R_0;t)\rho_0(R_\tau,R_0;\tau)<0$ for all $\tau<t\leq\beta$. We 
generally run our simulations with an acceptance ratio for the occupation of 
the Z sector close to 1/2. When calculating diagonal properties we consider 
the density matrix averaged over the entire path and not only at the reference 
point. For each move we can decide the frequency of the move and the maximum
number of time slices it operates on, apart from the displace move
where instead of the maximum number of time slices we can decide the
maximum extent of the spatial translation displacement. 

We noticed that doing like so, at low-temperature, the simulation with all the 
moves activated would enter the G sector without being able to get out of it (In 
order to exit the G sector the temporal distance between Ira and Masha must be 
close to 0 or $\beta$ and the spatial distance close to 0. The temporal distance 
is a stochastic variable which change of an amount $\beta$ in a number of moves of 
the order of $M^2$. So at larger $M$ the change of sector becomes rarer).
So at first we switched off the advance-recede and swap moves and more generally 
the access to the G sector (by properly adjusting the dimensionless parameter $C$ 
\cite{Boninsegni2006a,Boninsegni2006b} which controls the relative statistics
of Z and G-sectors) in our simulations. This is equivalent to restrict the 
configuration space to only the {\sl primal} nodal cell $\gamma_T(R_0)$ neglecting 
the other tiles obtained applying even permutations to the reference point $R_0$ 
according to the tiling property \cite{Ceperley1991}.  

In order to include correctly the permutations and the transition through the G 
sector of the worm algorithm, in our low temperature simulations, we had to use a 
different algorithm, that we will call algorithm B. Instead of using a generic G 
sector, we work in a restricted one where we impose equal imaginary times for Ira 
and Masha and a spatial distance between Ira and Masha equal to $\epsilon L$ with 
$\epsilon<1$ (here it is important not to take $\epsilon$ too small otherwise the 
acceptance ratios of the various moves ending in the G sector will go to zero). 
That is, rather than using the sector of the numerator of the whole Green’s 
function, one works with the sector of the single-particle density matrix 
at a distance less than $\epsilon L$. We accomplished 
this by constructing the following set of three, Z to G, G to Z, and G to G, 
moves obtained by combining the elementary moves of the usual worm algorithm 
\cite{Boninsegni2006a,Boninsegni2006b}: open-advance (removes a random number $m
$ of time slices and advances Ira of $m$ time slices), recede-close (recedes Ira 
by a random number $m$ of time slices and closes the worm), advance-recede 
(advances Ira by a random number $m$ of time slices and advances Masha by the 
same number of time slices). Moreover we just killed the usual insert and remove 
moves which would have to use a number of time slices equal to $M$ and would 
thus have very low acceptance ratios. Each of these three combined moves 
produces a configuration with an Ira and a Masha at the same imaginary time. 
We did not change all the other moves: swap, wiggle, and displace. This amounts 
to simulate a G sector for the one-body density matrix (which can be obtained 
from the histogram of the spatial distance between Ira and Masha). We note that 
this algorithm is inherently a canonical ensemble one. Moreover we rejected  
those moves which would bring to have a spatial distance between 
Ira and Masha larger than $\epsilon L$. We then introduced the 
nodal restriction also on this set of three moves: open-advance, recede-close, 
advance-recede, choosing as the reference point the one immediately next to Ira 
in imaginary time. 

We used this other algorithm to simulate just two of the low temperature cases 
among the twelve cases considered in the next section and observed a relevant 
improvement in the numerical results as compared with the existing literature 
data. This fact validated our algorithms.

It is well known that Monte Carlo algorithms works better as long as we have a
richer moves' menu, unless of course one violates detailed balance. So
our modified worm algorithm is very efficient in exploring all the electrons
path configurations with all the necessary permutation exchanges, even if in 
our restricted version the winding numbers will reflect the restriction. We 
will not be able to determine the superfluid fraction in our simulations. This 
is a shortcoming of applying the restricted path integral method where the winding 
numbers are biased by the restriction. 

%%%%%%%%%%%%%%%%%%%%%%%%%%%%%%%%%%%%%%%%%%%%%%%%%%%%%%%%%%%%%%%%%%%%%%%%%%%%%%
\section{Results} 
%%%%%%%%%%%%%%%%%%%%%%%%%%%%%%%%%%%%%%%%%%%%%%%%%%%%%%%%%%%%%%%%%%%%%%%%%%%%%%
\label{sec:results}

We simulated the Jellium at high density and low temperature. Given the bare
Coulomb potential $v(r)=2~\text{Ry}/r_sr$, according to
Fraser {\sl et al.} \cite{Fraser1996} it is possible to use in the simulation 
the following pair-potential $\phi$,
\bq \label{fraserpp}
\phi(r)&=&v(r)-\frac{N}{N-1}D,\\ \label{fraserD}
D&=&\frac{1}{\Omega}\int_{\rm cell}v(r)\,d\rr.
\eq
This method is equivalent to the Ewald summation technique or to its 
developments like the one carried on by Natoli and Ceperley \cite{Natoli1995} 
and gives smaller finite-size effects. The method is much more simple to 
implement than the more common Ewald sums but of course it has discontinuities 
when jumping from one side of the simulation cell to the other. The additive 
constant $D$ is chosen to make sure that the average value of the interaction is 
zero and the self energy of the electrons is taken as zero. 

In Table \ref{tab:tq} we present our results for various thermodynamic
quantities in the fully polarized $\xi=1$ case with $N=33$ particles. The 
statistical errors in the various measured quantities were determined, as usual, 
through the estimate of the correlation time of the given observable ${\cal O}$, $
\tau_{\cal O}$, as ${\rm error}=\sqrt{\tau_{\cal O}\sigma^2_{\cal O}/N}$ where $
\sigma^2_{\cal O}$ is the variance of ${\cal O}$ and $N$ is the number of MC steps. 
Our results can be directly compared with the ones of Brown {\sl et al.} 
\cite{Brown2013}. Benchmark data correcting systematic errors 
\cite{Ceperley1992} up to a 10\% in the high density $r_s\lesssim 1$ and low 
temperature cases of Brown {\sl et al.} can be found 
in Refs. \cite{Schoof2015,Dornheim2016,Dornheim2016b,Groth2016,Groth2017}. 
The time steps $\tau$ chosen in the simulations are like the ones chosen by 
Brown {\sl et al.} \cite{Brown2013} as a function of $r_s$ at all temperatures: 
$\tau=0.0007$ for $r_s=1$, $\tau=0.0027$ for $r_s=2$, and $\tau=0.0214$ for 
$r_s=4$ but in any case with $M$ not bigger than $10^3$. From the table we can 
see how our results agree well with the ones of Brown {\sl et al.} 
\cite{Brown2013}: The kinetic energy, in the highest density case, is within a 
0.5\% at high temperatures (in the correct direction given by the later results 
of Refs. \cite{Schoof2015,Groth2016}) and up to a 35\% in the lower temperature 
case. This discrepancy increase is due to the fact that in these 
simulations we had the advance-recede and swap moves switched off, so we were 
not sampling the whole fermions configuration space but only the primal nodal 
cell (the one connected directly to the reference point itself), as explained in 
the previous section. This clearly becomes more and more important 
at low temperature when the quantum effects are more relevant. 

\begin{table}[htbp]
\caption{Thermodynamic results in our simulations with $\xi=1$ and $N=33$ 
  electrons interacting through the pair-potential $\phi(r)=v(r)-ND/(N-1)$ of 
  Eqs. (\ref{fraserpp})-(\ref{fraserD}), at a density fixed by $r_s$, 
  temperature fixed by $\Theta$ (at a Coulomb coupling constant $\Gamma$), and 
  with $M$ time slices: $e_0~(\text{Ry})$ is the internal energy per particle 
  of the ideal gas, $P_0~(\text{Ry}/r_s^3a_0^3)$ is the pressure of the ideal 
  gas, $e_k~(\text{Ry})$ is the kinetic energy per particle in our simulation,
  $e_k^{\rm Brown}~(\text{Ry})$ is the kinetic energy per particle in Brown {\sl   
  et al.} \cite{Brown2013} simulation,
  $e_p~(\text{Ry})$ is the potential energy per particle in our simulation,
  $e_p^{\rm Brown}~(\text{Ry})$ is the potential energy per particle in Brown   	
  {\sl et al.} \cite{Brown2013} simulation, 
  $e_t~(\text{Ry})=e_k+e_p$ is the total energy per particle in our simulation, 
  and $P~(\text{Ry}/r_s^3a_0^3)$ is the pressure in our simulation. In these   
  simulations we used algorithm A with the advance-recede and swap moves 
  switched off.} 
\label{tab:tq}
\vspace{.5cm}
%\begin{ruledtabular}
{\footnotesize \begin{tabular}{||c|ccc|cc|cc|cc|cc||}
\hline
\hline
$M$ & $r_s$ & $\Theta$ & $\Gamma$ & $e_0$ & $P_0$ & $e_k^{\rm Brown}$ & $e_p^{\rm Brown}$ & $e_k$ & $e_p$ & $e_t$ & $P$ \\
\hline 
244 & 1 & 1 & 0.342 & 9.920268 & 1.578860 & 9.72(2) & $-$0.938(1) & 9.67(5) & $-$0.970(3) & 8.70(5) & 2.670(7) \\
489 & 1 & 0.5 & 0.684 & 5.973201 & 0.950664 & 5.72(2)& $-$1.088(1) & 5.67(8) & $-$1.133(3) & 4.53(8) & 2.02(1) \\
977 & 1 & 0.25 & 1.368 & 4.307310 & 0.685530 & 4.12(4) & $-$1.171(1) & 4.9(1) &  $-$1.233(2) & 3.7(1) & 1.89(2) \\
1000 & 1 & 0.125 & 2.737 & 3.727579 & 0.593263 & 3.64(1) & $-$1.1961(5) & 4.73(6) & $-$1.276(1) & 3.46(6) & 1.861(9) \\
\hline
253 & 2 & 1 & 0.684 & 2.480067 & 0.394715 & 2.419(5) & $-$0.5280(4) & 2.39(1) & $-$0.542(1) & 1.85(1) & 0.941(2) \\
507 & 2 & 0.5 & 1.368 & 1.493300 & 0.237666 & 1.435(5) & $-$0.5917(2) & 1.46(2) & $-$0.612(1) & 0.85(2) & 0.788(3) \\
1000 & 2 & 0.25 & 2.737 & 1.076827 & 0.171382 & 1.050(7) & $-$0.6219(2) & 1.24(3) & $-$0.6484(9) & 0.59(3) & 0.750(4) \\
1000 & 2 & 0.125 & 5.473 & 0.931895 & 0.148316 & 0.906(4) & $-$0.6302(1) & 1.22(2) & $-$0.663(1) & 0.55(2) & 0.745(4) \\
\hline
128 & 4 & 1 & 1.368 & 0.620017 & 0.098679 & 0.597(1) & $-$0.2885(3)* & 0.593(1) & $-$0.3026(1) & 0.290(1) & 0.3725(2) \\
256 & 4 & 0.5 & 2.737 & 0.373325 & 0.059416 & 0.367(1) & $-$0.3206(1) & 0.361(2) & $-$0.3282(2) & 0.033(2) & 0.3335(3) \\
512 & 4 & 0.25 & 5.473 & 0.269207 & 0.042846 & 0.269(1) & $-$0.3302(1) & 0.303(2) & $-$0.3396(1) & $-$0.036(2) & 0.3234(3) \\
1000 & 4 & 0.125 & 10.946 & 0.232974 & 0.037079 & 0.237(1) & $-$0.3318(1) & 0.30(1) & $-$0.3444(6) & $-$0.05(1) & 0.322(2) \\
\hline
\hline
\end{tabular}}
%\end{ruledtabular}
\end{table}

The data denoted with an asterisk in the table has been considerably corrected 
by the later work of Groth {\sl et al.} \cite{Groth2016}, who give 
$e_p=-0.305012(33)$, which is much closer to our result.

\begin{figure}[htbp]
\begin{center}
\includegraphics[width=10cm]{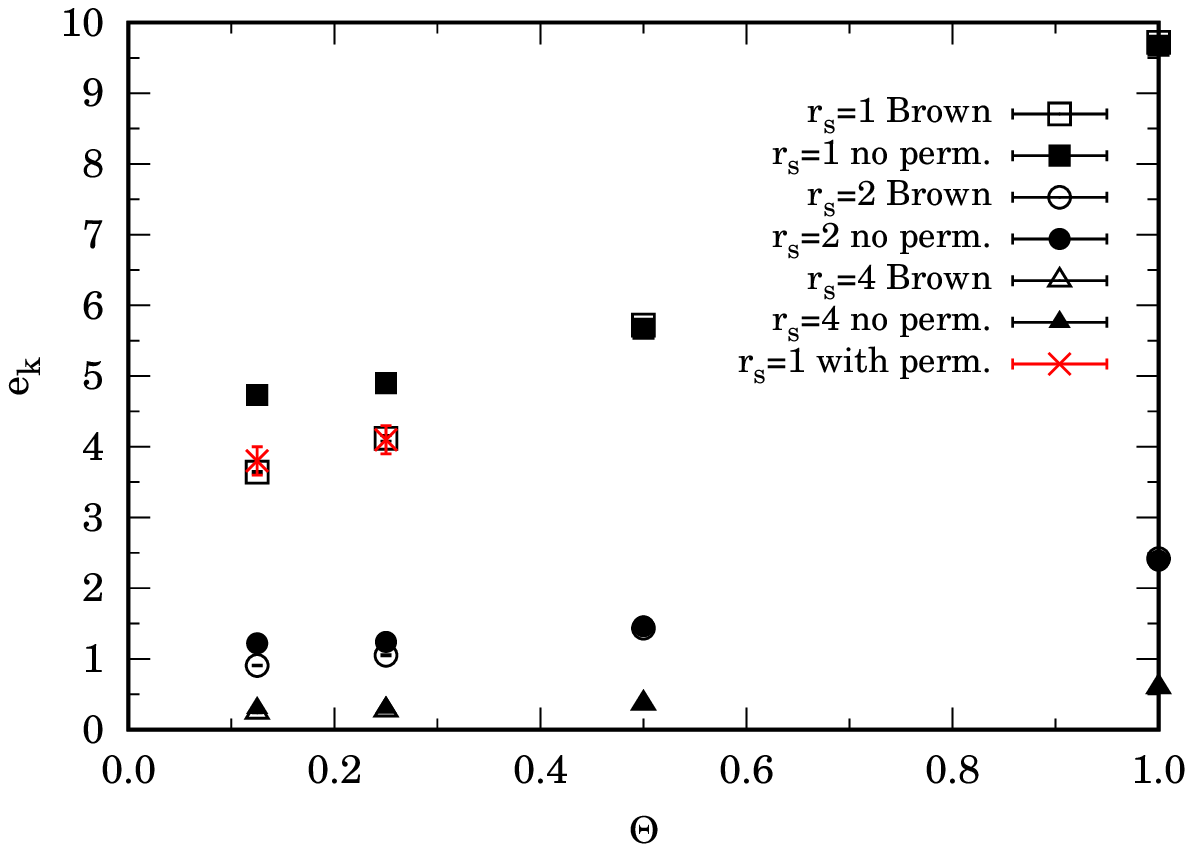}\\
\includegraphics[width=10cm]{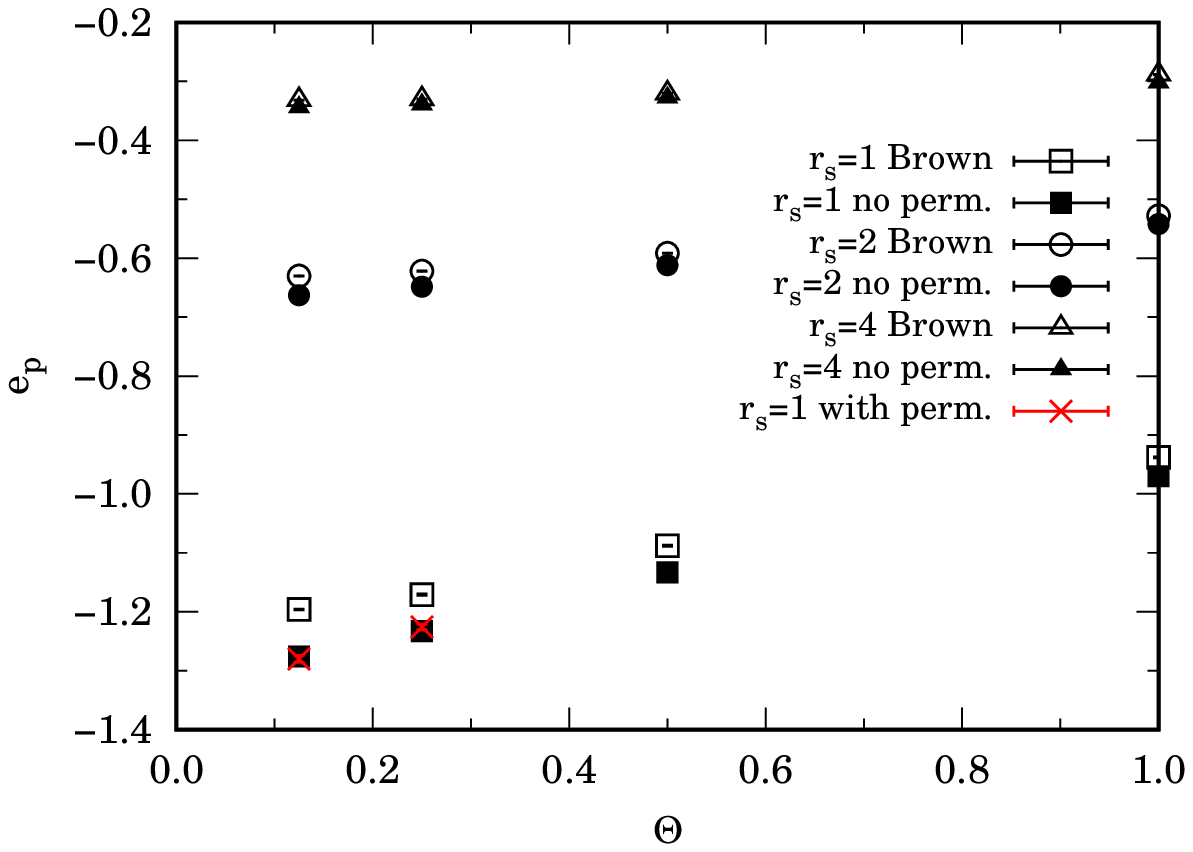}
\end{center}  
\caption{(color online) We show a comparison of our results at the three different 
values of density ($r_s=1,2,4$), with (with perm.) and 
without (no perm.) permutations, for the kinetic energy per particle (top panel) 
and the potential energy per particle (bottom panel) with the results of Brown {\sl 
et al.} \cite{Brown2013} (Brown) as they are reported in the Tables \ref{tab:tq} 
(no perm.) and \ref{tab:tqB} (with perm.).}
\label{fig:compare}
\end{figure}

In Fig. \ref{fig:compare} we show a comparison of our results for the kinetic 
energy per particle (top panel) and the potential energy per particle (bottom 
panel) with the results of Brown {\sl et al.} \cite{Brown2013}. From the Figure we 
see clearly how our results with no permutations reproduce well the results of 
Brown {\sl et al.} at sufficiently high temperatures and low densities. And our 
results with the permutations switched on corrects the discrepancy observed at low 
temperatures (small $\Theta$) and high densities (small $r_s$).

\begin{figure}[htbp]
\begin{center}
\includegraphics[width=10cm]{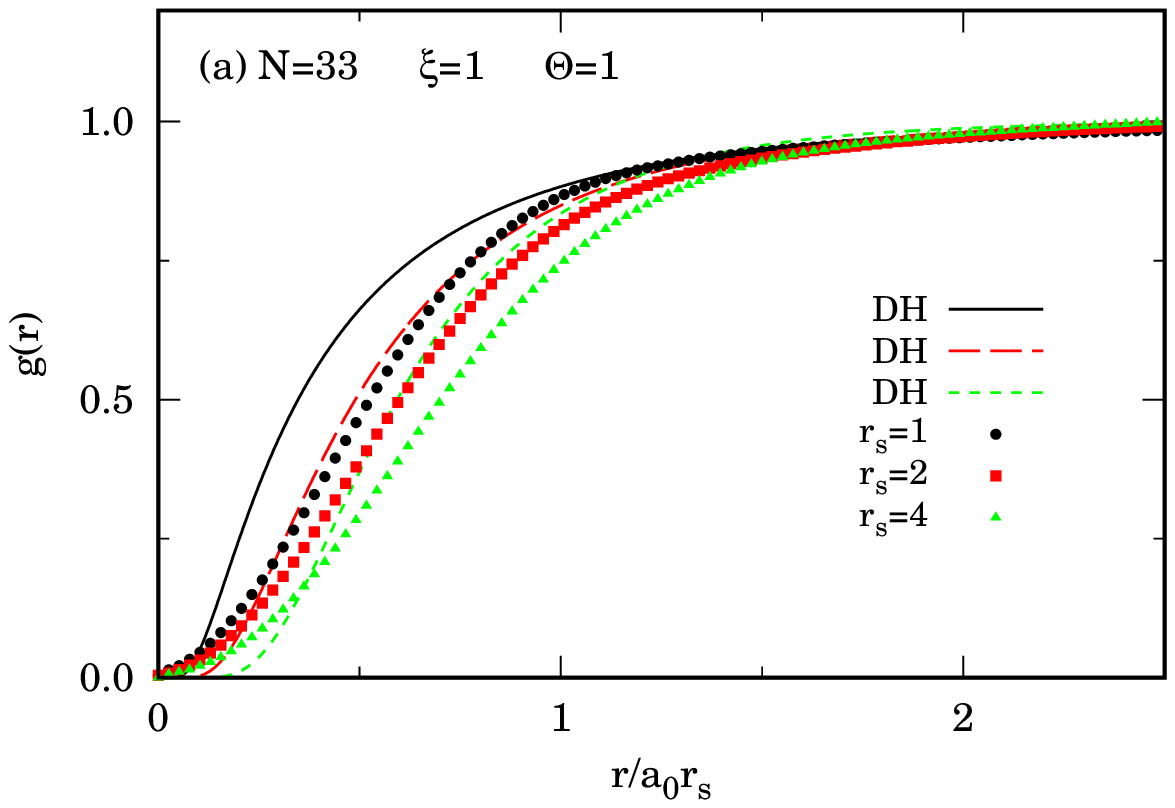}\\
\includegraphics[width=10cm]{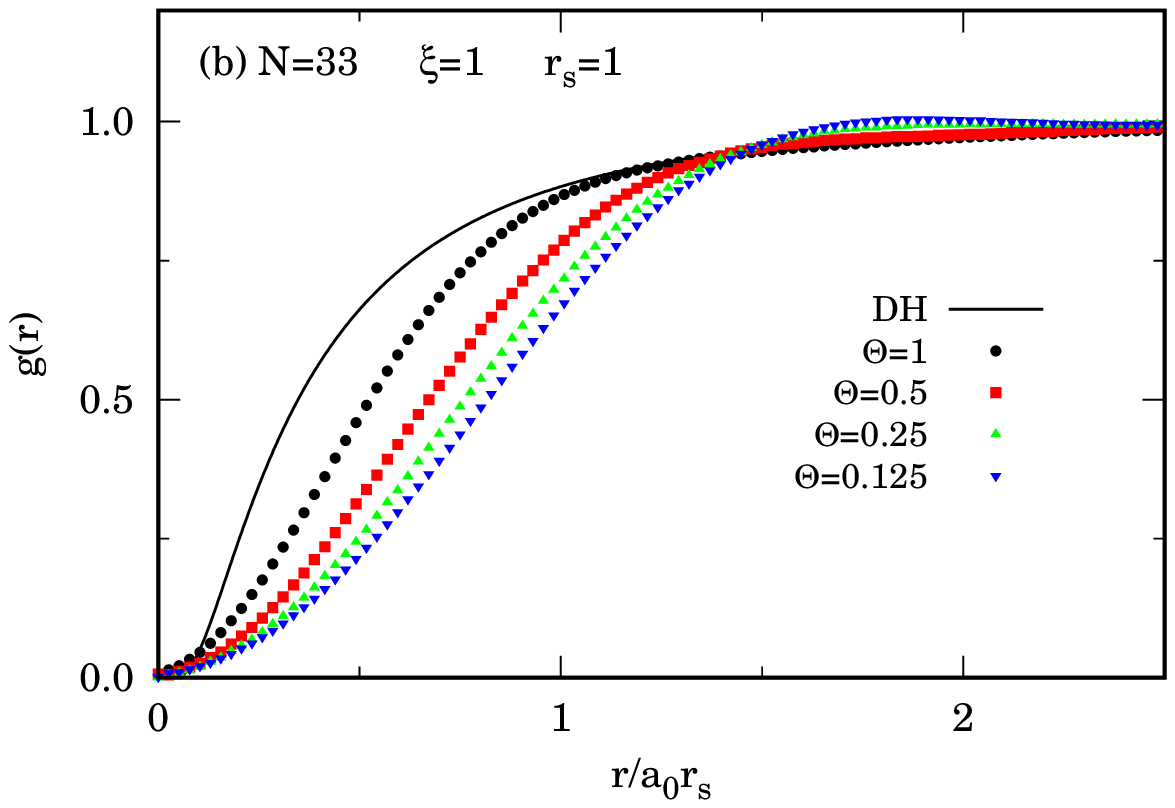}
\end{center}  
\caption{(color online) The radial distribution function for Jellium
  in selected states of Table \ref{tab:tq}, from algorithm A, at fixed 
  temperature in the upper 
  panel (a) and at fixed density in the lower panel (b). Also shown is the 
  Debye-H\"uckel (DH) result \cite{Martin88} for the high temperature and low   	
  density limit, $g_{DH}(r)=\exp\left[-\frac{\Gamma}{r}
  \exp\left(-\sqrt{3\Gamma}r\right)\right]$.}
\label{fig:gofr}
\end{figure}

In Fig. \ref{fig:gofr} we show our results for the radial distribution function 
\cite{Allen-Tildesley}, $g(r)$, for selected states of Table \ref{tab:tq} at 
fixed temperature and at fixed density, respectively.

As outlined in the previous section we repeated the calculation for the low 
temperature cases $\xi=1, r_s=1, \Theta=0.25$ and $\Theta=0.125$ with our 
modified algorithm B, with $\epsilon=1/2$, able to sample the whole fermions 
configuration space including the necessary permutations. The result in these 
cases were encouraging and are shown in Table \ref{tab:tqB}. They were much closer 
to the corresponding result of Brown {\sl et al.} \cite{Brown2013} than the 
results obtained with the previous algorithm A: The kinetic energy, in the highest 
density case, is within a 5\% at low temperatures. We also checked that the two 
algorithms, A and B, coincide at high temperature. This validates our algorithms 
A and B.

\begin{table}[htbp]
\caption{Same as Table \ref{tab:tq} but using our algorithm B in the high density 
low temperature simulations.} 
\label{tab:tqB}
\vspace{.5cm}
%\begin{ruledtabular}
{\footnotesize \begin{tabular}{||c|ccc|cc|cc|cc|cc||}
\hline
\hline
$M$ & $r_s$ & $\Theta$ & $\Gamma$ & $e_0$ & $P_0$ & $e_k^{\rm Brown}$ & $e_p^{\rm Brown}$ & $e_k$ & $e_p$ & $e_t$ & $P$ \\
\hline 
977 & 1 & 0.25 & 1.368 & 4.307310 & 0.685530 & 4.12(4) & $-$1.171(1) & 4.1(2) &  $-$1.226(5) & 2.9(2) & 1.76(4) \\
1000 & 1 & 0.125 & 2.737 & 3.727579 & 0.593263 & 3.64(1) & $-$1.1961(5) & 3.8(2) & $-$1.280(6) & 2.5(2) & 1.73(2) \\
\hline
\hline
\end{tabular}}
%\end{ruledtabular}
\end{table}

\begin{figure}[htbp]
\begin{center}
\includegraphics[width=10cm]{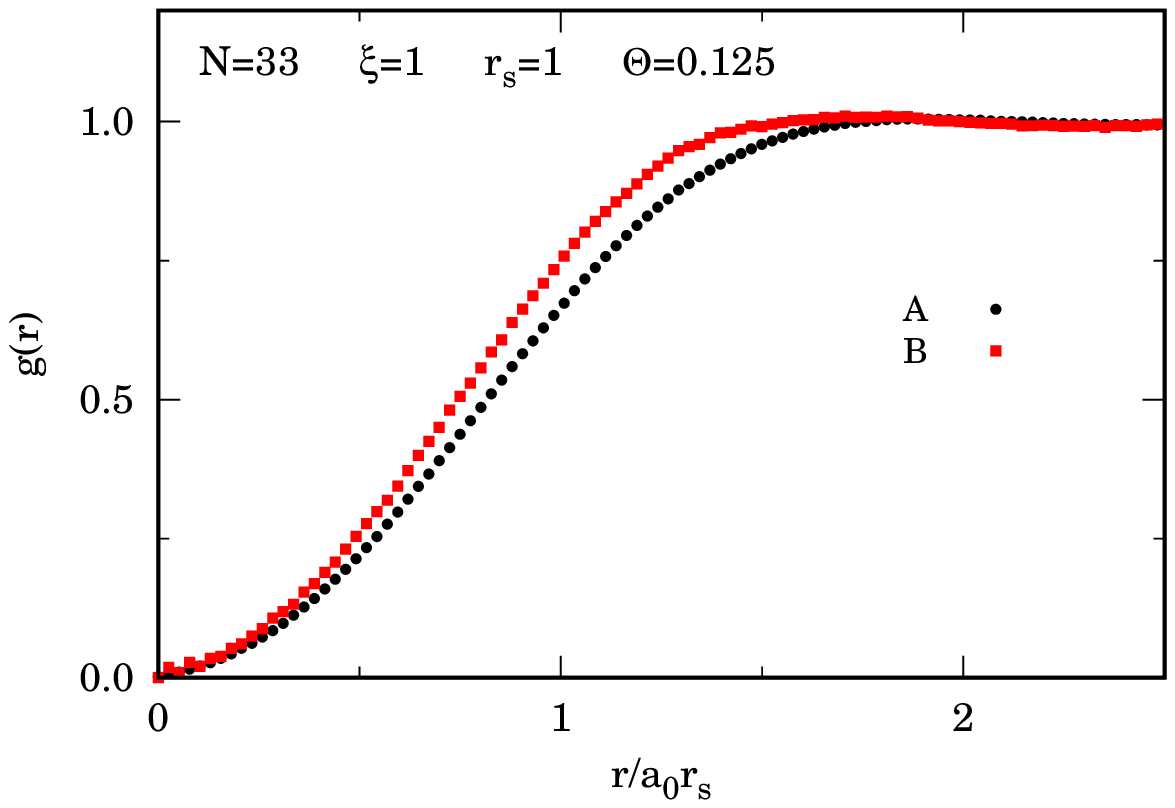}
\end{center}  
\caption{(color online) The radial distribution function for Jellium
  in the $\xi=1, r_s=1, \Theta=0.125$ state as obtained from    
  our two algorithms A and B: The one without G sector and the one with 
  G sector, respectively.}
\label{fig:gofrlt}
\end{figure}

In Fig. \ref{fig:gofrlt} we show our results for the radial distribution 
function for the $\xi=1, r_s=1, \Theta=0.125$ state obtained with the algorithm 
with the G sector switched off (A) and with the algorithm with the 
G sector switched on (B).
From the figure we see how the Fermi hole diminishes by the introduction of the 
permutations in the calculation.
%%%%%%%%%%%%%%%%%%%%%%%%%%%%%%%%%%%%%%%%%%%%%%%%%%%%%%%%%%%%%%%%%%%%%%%%%%%%%%
\section{Conclusions} 
%%%%%%%%%%%%%%%%%%%%%%%%%%%%%%%%%%%%%%%%%%%%%%%%%%%%%%%%%%%%%%%%%%%%%%%%%%%%%%
\label{sec:conclusions}

We have successfully implemented the ideal fermion density matrix
restriction on the path integral worm algorithm which is able to
generate the necessary RPIMC moves during the simulation evolution
thereby circumventing the otherwise inevitable sign problem. This allowed 
us to reach the finite, non-zero, temperature properties of a given fluid model 
of Fermi particles interacting through a given pair-potential. We worked in the 
canonical ensemble and applied our method to the Jellium fluid of
Wigner. We explicitly compared our results with the previous canonical 
calculation of Brown {\sl et al.} \cite{Brown2013} in the high density and low 
temperature regime where their algorithm had problems in sampling the path 
\cite{Ceperley1992}.
Our results complement the ones of Brown {\sl et al.} with 
the treatment of the high density $r_s\le 4$ and low temperature cases which 
were found to be inaccurate by Bonitz {\sl et al.}  
\cite{Schoof2015,Groth2016,Groth2017} who suggested an alternative algorithm to 
circumvent the systematic errors in Brown calculations \cite{Ceperley1992}. 

The relevance of our study relies in the fact that our simulation method is 
different from both the method of Ceperley {\sl et al.} 
\cite{Brown2013,Brown2014} who uses the fixed-nodes approximation in the 
canonical ensemble of a regular, and not worm, PIMC \cite{Ceperley1995}, 
and from the one of Bonitz {\sl et al.} 
\cite{Schoof2011,Dornheim2015,Dornheim2016,Groth2017} who combine configuration- 
and permutation-blocking PIMC. Our method is also different from
other quantum Monte Carlo methods like the one of Malone {\sl et al.}
\cite{Malone2016} that agrees well with the one of Bonitz {\sl et al.} at high
densities and the direct PIMC one of Filinov {\sl et al.} \cite{Filinov2015} that 
agrees well with Brown {\sl et al.} at low density and moderate temperature. So 
our new algorithms add to the ones already used in the quest for an optimal way to 
calculate the properties of the fascinating Wigner's Jellium model at finite, non 
zero, temperatures. We devised two different algorithms, A and B. In algorithm A 
we used a restricted, fixed-nodes, worm algorithm which never passes through the G 
sector. In algorithm B we used a restricted, fixed-nodes, worm algorithm with a G 
sector which has Masha and Ira always at the same imaginary time and at a given 
small spatial distance. In both cases the restriction of the fixed-nodes path 
integral is the one from a trial density matrix equal to the one of ideal 
fermions.

We obtained results for both the static structure (the radial distribution 
function) and various thermodynamic quantities (energy and pressure) for the 
Jellium model with $N=33$ fully polarized ($\xi=1$) electrons at high density 
and low temperature. Our results compares favorably with the ones of Brown {\sl  
et al.} \cite{Brown2013} with a discrepancy on the kinetic energy, in the 
highest density case, up to a 0.5\% at high temperatures (with our algorithm A) 
and up to 5\% at low temperatures (with our algorithm B). Our results can also be 
compared with the later ones of Refs. \cite{Schoof2015,Groth2016} with which the 
agreement increases even further. This validates our algorithms which are 
alternative to the ones that have already been used in the literature.

We expect in the near future to explicitly determine the dependence of the 
Jellium properties (structural and thermodynamic) on the polarization $\xi$.
We would also like to carry out a more comprehensive comparison with the results 
in the literature and to predict other results yet to be determined through 
quantum Monte Carlo methods, like the static structure function. Regarding 
improvements to the algorithm we would like to implement the use of better
approximations for the action in the path integral and a search for
better trial density matrices to guide the fixed-nodes at low
temperatures or the implementation of the released-nodes recipe.

Another important problem to solve is the one of calculating the superfluid 
fraction for fermions or superconducting fraction for electrons. The winding 
numbers that one is computing in RPIMC are not be sufficient to determine the 
superfluid fraction since there is the restriction on the paths.

%%%%%%%%%%%%%%%%%%%%%%%%%%%%%%%%%%%%%%%%%%%%%%%%%%%%%%%%%%%%%%%%%%%%%%%%%%%%%% 
\begin{acknowledgments}
We would like to thank Saverio Moroni for several relevant discussions at 
S.I.S.S.A. of Trieste, Boris Svistunov for useful e-mail and Skype suggestions on 
how to implement our algorithm B, and David Ceperley for many e-mail exchanges
which have been determinant for the completion of the work. 
\end{acknowledgments} 
%%%%%%%%%%%%%%%%%%%%%%%%%%%%%%%%%%%%%%%%%%%%%%%%%%%%%%%%%%%%%%%%%%%%%%%%%%%%%%
%\bibliographystyle{}
%\bibliography{jft}


%%%%%%%%%%%%%%%%%%%%%%%%%%%%%%%%%%%%%%%%%%%%%%%%%%%%%%%%%%%%%%%%%%%%%%%%%%%%%%
%%%%%%%%%%%%%%%%%%%%%%%%%%%%%%%%%%%%%%%%%%%%%%%%%%%%%%%%%%%%%%%%%%%%%%%%%%%%%%
%%%%%%%%%%%%%%%%%%%%%%%%%%%%%%%%%%%%%%%%%%%%%%%%%%%%%%%%%%%%%%%%%%%%%%%%%%%%%%
\end{document}